\documentclass[11pt,a4paper]{article}

\usepackage[utf8]{inputenc}
\usepackage[T1]{fontenc}
\usepackage{lmodern}
\usepackage[margin=1in]{geometry}
\usepackage{graphicx}
\usepackage{hyperref}
\usepackage{amsmath,amssymb}
\usepackage{booktabs}
\usepackage{array}
\usepackage{tabularx}
\usepackage{listings}
\usepackage{xcolor}
\usepackage{float}
\usepackage{caption}
\usepackage{subcaption}
\usepackage{enumitem}
\usepackage{tikz}
\usetikzlibrary{shapes.geometric, arrows, positioning, fit, backgrounds}
\usepackage{pgfplots}
\pgfplotsset{compat=1.18}
\usepackage{pgf-pie}
\usepackage{natbib}
\usepackage{url}
\usepackage{microtype}

\hypersetup{
    colorlinks=true,
    linkcolor=blue,
    filecolor=magenta,
    urlcolor=cyan,
    citecolor=blue,
    pdftitle={Constitutional Spec-Driven Development: Enforcing Security by Construction in AI-Assisted Code Generation},
    pdfauthor={Srinivas Rao Marri},
}

\definecolor{codegreen}{rgb}{0,0.6,0}
\definecolor{codegray}{rgb}{0.5,0.5,0.5}
\definecolor{codepurple}{rgb}{0.58,0,0.82}
\definecolor{backcolour}{rgb}{0.95,0.95,0.92}

\lstdefinestyle{pythonstyle}{
    backgroundcolor=\color{backcolour},
    commentstyle=\color{codegreen},
    keywordstyle=\color{magenta},
    numberstyle=\tiny\color{codegray},
    stringstyle=\color{codepurple},
    basicstyle=\ttfamily\footnotesize,
    breakatwhitespace=false,
    breaklines=true,
    captionpos=b,
    keepspaces=true,
    numbers=left,
    numbersep=5pt,
    showspaces=false,
    showstringspaces=false,
    showtabs=false,
    tabsize=2,
    language=Python
}

\tikzstyle{constitution} = [rectangle, rounded corners, minimum width=3cm, minimum height=1cm, text centered, draw=black, fill=red!30, text=black]
\tikzstyle{speclayer} = [rectangle, rounded corners, minimum width=3cm, minimum height=1cm, text centered, draw=black, fill=blue!30, text=black]
\tikzstyle{ailayer} = [rectangle, rounded corners, minimum width=3cm, minimum height=1cm, text centered, draw=black, fill=purple!30, text=black]
\tikzstyle{implementation} = [rectangle, rounded corners, minimum width=3cm, minimum height=1cm, text centered, draw=black, fill=green!30, text=black]
\tikzstyle{traceability} = [rectangle, rounded corners, minimum width=3cm, minimum height=1cm, text centered, draw=black, fill=orange!30, text=black]
\tikzstyle{arrow} = [thick,->,>=stealth]

\title{
    \textbf{Constitutional Spec-Driven Development:\\Enforcing Security by Construction in AI-Assisted Code Generation}
}

\author{
    Srinivas Rao Marri\\
    \texttt{srinivasrao.marri@gmail.com}\\
    \\
    \small{January 2026}
}

\date{}

\begin{document}

\maketitle

% ============================================================
% ABSTRACT
% ============================================================
\begin{abstract}
The proliferation of AI-assisted ``vibe coding'' enables rapid software development but introduces significant security risks, as Large Language Models (LLMs) prioritize functional correctness over security. We present \textit{Constitutional Spec-Driven Development}, a methodology that embeds non-negotiable security principles into the specification layer, ensuring AI-generated code adheres to security requirements by construction rather than inspection. Our approach introduces a \textit{Constitution}: a versioned, machine-readable document encoding security constraints derived from Common Weakness Enumeration (CWE)/MITRE Top 25 vulnerabilities and regulatory frameworks. We demonstrate the methodology through a banking microservices application, selected as a representative example domain due to its stringent regulatory and security requirements, implementing customer management, account operations, and transaction processing. The methodology itself is domain-agnostic. The implementation addresses 10 critical CWE vulnerabilities through constitutional constraints with full traceability from principles to code locations. Our case study shows that constitutional constraints reduce security defects by 73\% compared to unconstrained AI generation while maintaining developer velocity. We contribute a formal framework for constitutional security, a complete development methodology, and empirical evidence that proactive security specification outperforms reactive security verification in AI-assisted development workflows.

\vspace{0.5em}
\end{abstract}

% ============================================================
% 1. INTRODUCTION
% ============================================================
\section{Introduction}
\label{sec:introduction}

The rapid adoption of AI-assisted code generation has created a fundamental tension in software development: the same tools that dramatically accelerate development velocity also introduce systematic security vulnerabilities. This paper addresses this challenge by introducing the term \textit{Constitutional Spec-Driven Development (CSDD)}, a novel methodology that embeds non-negotiable security constraints into the specification layer, ensuring that AI-generated code is secure by construction rather than by post-hoc verification. To our knowledge, this represents the first formalization of constitutional constraints applied specifically to AI code generation workflows, combining principles from Design by Contract~\cite{meyer1992applying}, Constitutional AI~\cite{bai2022constitutional}, and specification-driven development into a unified framework for secure AI-assisted software engineering. We use the term ``constitution'' metaphorically: just as a political constitution establishes foundational principles that govern all subsequent legislation, a software constitution establishes non-negotiable constraints that govern all subsequent code generation. The term implies architectural primacy, not legal obligation.

\subsection{Motivation}

\textbf{The Rise of Vibe Coding.} The emergence of AI-assisted coding tools has fundamentally transformed software development. Large Language Models can generate functional code from natural language descriptions, enabling rapid prototyping and reducing implementation time. This paradigm, termed ``vibe coding,'' allows developers to describe desired functionality conversationally and receive working implementations. A developer might simply state ``create a user registration endpoint'' and receive a complete implementation within seconds. However, this acceleration introduces significant security risks that traditional development practices are ill-equipped to address.

\vspace{0.5em}
\textbf{The Security Gap in AI-Generated Code.} Studies indicate that LLM-generated code frequently contains security vulnerabilities~\cite{pearce2022asleep, perry2023users, zhou2025vibecoding}. Analysis of code produced by popular AI assistants reveals patterns of SQL injection, cross-site scripting, improper authentication, and insufficient input validation, vulnerabilities catalogued in the CWE/MITRE Top 25 Most Dangerous Software Weaknesses~\cite{cwe2025}. The fundamental issue is that AI models optimize for functional correctness based on training data distributions, not security requirements specific to deployment contexts. When an AI generates a database query, it produces code that \textit{works}, but ``working'' code that concatenates user input into SQL strings creates exploitable injection vulnerabilities.

\vspace{0.5em}
\textbf{Regulatory and Industry Pressures.} The banking and financial services sector presents acute challenges. Regulatory frameworks including PCI-DSS, SOC 2, and GDPR impose strict security requirements~\cite{pcidss}. A single SQL injection vulnerability in a banking application could expose millions of customer records, resulting in regulatory fines exceeding \$10 million, class-action litigation, and irreparable reputation damage. Traditional security practices (code reviews, penetration testing, static analysis) operate as post-hoc verification, detecting vulnerabilities after introduction rather than preventing creation. When AI accelerates code generation by orders of magnitude, reactive security processes become inadequate.

\vspace{0.5em}
\textbf{The False Dichotomy.} The tension appears fundamental: organizations desire AI-assisted velocity while requiring security guarantees mandated by regulation. Current approaches treat these as competing objectives; teams must choose between moving fast with AI or moving carefully with security. We argue this dichotomy is false and propose a methodology achieving both through architectural intervention at the specification layer. By constraining AI generation \textit{before} code is produced, rather than inspecting it \textit{after}, we eliminate entire categories of vulnerabilities while preserving the velocity benefits of AI assistance.

\vspace{0.5em}
\textbf{The Core Problem.} Contemporary AI coding assistants operate without persistent security constraints. Each request is processed independently, relying on prompt engineering to incorporate security considerations. This suffers from:

\begin{itemize}[noitemsep,topsep=3pt]
    \item \textbf{Inconsistency}: Security requirements must be restated per prompt, and developers frequently forget to include critical constraints like ``use parameterized queries'' or ``hash passwords with bcrypt''
    \item \textbf{Incompleteness}: Developers omit requirements they consider obvious, but AI models lack the contextual understanding to infer that a banking application requires stronger security than a prototype
    \item \textbf{Drift}: Early security specifications do not propagate to later code. A developer who specified secure authentication in sprint one may forget to maintain those standards when implementing new features in sprint five
    \item \textbf{Unverifiability}: No systematic mechanism exists to verify that generated code adheres to stated requirements across an entire codebase
\end{itemize}

These limitations are dangerous in vibe coding workflows where developers accept generated implementations with minimal scrutiny. The cognitive offloading that makes AI assistance attractive simultaneously reduces security flaw detection likelihood.

\subsection{Contributions}

This paper introduces Constitutional Spec-Driven Development as a new paradigm and makes the following contributions:
\vspace{0.3em}

\textbf{C1. Constitutional Security Framework.} We formalize software constitutions, hierarchical constraint systems encoding non-negotiable security requirements as first-class architectural artifacts with versioning and governance. Unlike ad-hoc security guidelines, constitutions are structured documents with explicit CWE mappings, enforcement levels (MUST/SHOULD/MAY), and amendment procedures.

\vspace{0.4em}
\textbf{C2. Spec-Driven Development Methodology.} We present a complete workflow integrating constitutional constraints with AI-assisted code generation across specification, planning, task decomposition, and implementation phases.

\vspace{0.4em}
\textbf{C3. Compliance Traceability Matrix.} We introduce systematic mapping of constitutional principles to implementation artifacts at file and line-number granularity, enabling automated compliance verification.

\vspace{0.4em}
\textbf{C4. Reference Implementation.} We demonstrate the methodology through a banking microservices application addressing 10 CWE/MITRE Top 25 vulnerabilities through constitutional constraints.

\vspace{0.4em}
\textbf{C5. Empirical Evaluation.} We analyze constitutional constraint effectiveness, measuring compliance rates and defect density compared to unconstrained generation. Our case study demonstrates a 73\% reduction in security vulnerabilities, 56\% faster time to first secure build, and 4.3x improvement in compliance documentation coverage.

% ============================================================
% 2. BACKGROUND
% ============================================================
\section{Background and Context}
\label{sec:background}

This section reviews the foundational concepts underlying Constitutional Spec-Driven Development: AI-assisted code generation, known security challenges, and prior work in formal specification and constitutional AI.

\subsection{AI-Assisted Code Generation}

Large Language Models trained on code corpora can generate syntactically correct, functionally appropriate code from natural language prompts~\cite{chen2021codex}. Tools like GitHub Copilot, Claude, and ChatGPT have achieved widespread adoption, with surveys indicating over 70\% of professional developers use AI assistance regularly~\cite{github2024octoverse}.

\subsection{Security Challenges in AI-Generated Code}

Research has documented systematic security issues in AI-generated code. Pearce et al.~\cite{pearce2022asleep} found that GitHub Copilot produces vulnerable code in approximately 40\% of security-relevant scenarios. Perry et al.~\cite{perry2023users} demonstrated that developers using AI assistance write less secure code than those coding manually, partially due to over-reliance on AI output correctness. Zhou et al.~\cite{zhou2025vibecoding} specifically benchmarked vibe coding vulnerabilities in real-world tasks, confirming that agent-generated code exhibits systematic security weaknesses across diverse application domains. More recently, Liu et al.~\cite{liu2026agentskills} conducted a large-scale empirical study of 31,132 AI agent skills, finding that 26.1\% contain at least one security vulnerability, including prompt injection, data exfiltration, and privilege escalation risks. Their findings underscore that the specification artifacts guiding AI agents (including constitutional documents) must themselves be authored defensively to resist adversarial manipulation.

\subsection{Design by Contract}

Meyer's Design by Contract~\cite{meyer1992applying, meyer1997object} introduced formal specification of software behavior through preconditions, postconditions, and invariants. Our constitutional approach extends this tradition by applying contract-like constraints to AI code generation.

\subsection{Constitutional AI}

Anthropic's Constitutional AI~\cite{bai2022constitutional} demonstrated that AI systems can be guided by explicit principles during training and inference. We adapt this concept from AI alignment to software security, embedding security principles as generation constraints.

% ============================================================
% 3. METHODOLOGY
% ============================================================
\section{Constitutional Spec-Driven Development}
\label{sec:methodology}

This section presents the core methodology, introducing the key concepts, architectural components, and workflow that comprise Constitutional Spec-Driven Development.

\subsection{Core Concepts}

\textbf{Constitution.} By analogy to political constitutions that establish foundational governing principles, we define a software constitution as a versioned document encoding non-negotiable requirements as machine-readable principles with explicit CWE vulnerability mappings, enforcement levels (MUST/SHOULD/MAY per RFC 2119), and rationale. A constitution is not prescriptive about implementation technology; it specifies \textit{what} must hold, not \textit{how} to achieve it.

\vspace{0.4em}
\textbf{Spec-Driven Development.} A hierarchical workflow where constitutional constraints flow through specification, planning, and implementation phases, ensuring security requirements propagate to generated code.

\vspace{0.4em}
\textbf{Compliance Traceability.} Systematic mapping from constitutional principles to code artifacts at file and line-number granularity, enabling audit and impact analysis.

\subsection{Constitutional Principles Structure}

Each principle in our banking constitution follows this structure:

\begin{enumerate}[noitemsep,topsep=3pt]
    \item \textbf{Identifier}: Unique code (e.g., SEC-002)
    \item \textbf{CWE Reference}: Specific vulnerability addressed (e.g., CWE-89)
    \item \textbf{Enforcement Level}: MUST, SHOULD, or MAY
    \item \textbf{Constraint}: What the code must/must not do
    \item \textbf{Implementation Pattern}: How to satisfy the constraint
    \item \textbf{Rationale}: Why this constraint exists (attack vector)
\end{enumerate}

\subsection{Compliance Traceability Matrix}

The matrix maps each principle to specific implementation artifacts:

\begin{itemize}[noitemsep,topsep=3pt]
    \item \textbf{Audit Support}: Demonstrable compliance for regulators
    \item \textbf{Change Impact Analysis}: Understanding which code affects which principles
    \item \textbf{Gap Detection}: Identifying unimplemented requirements
    \item \textbf{Regression Prevention}: Ensuring changes do not violate principles
\end{itemize}

\subsection{Architecture Overview}

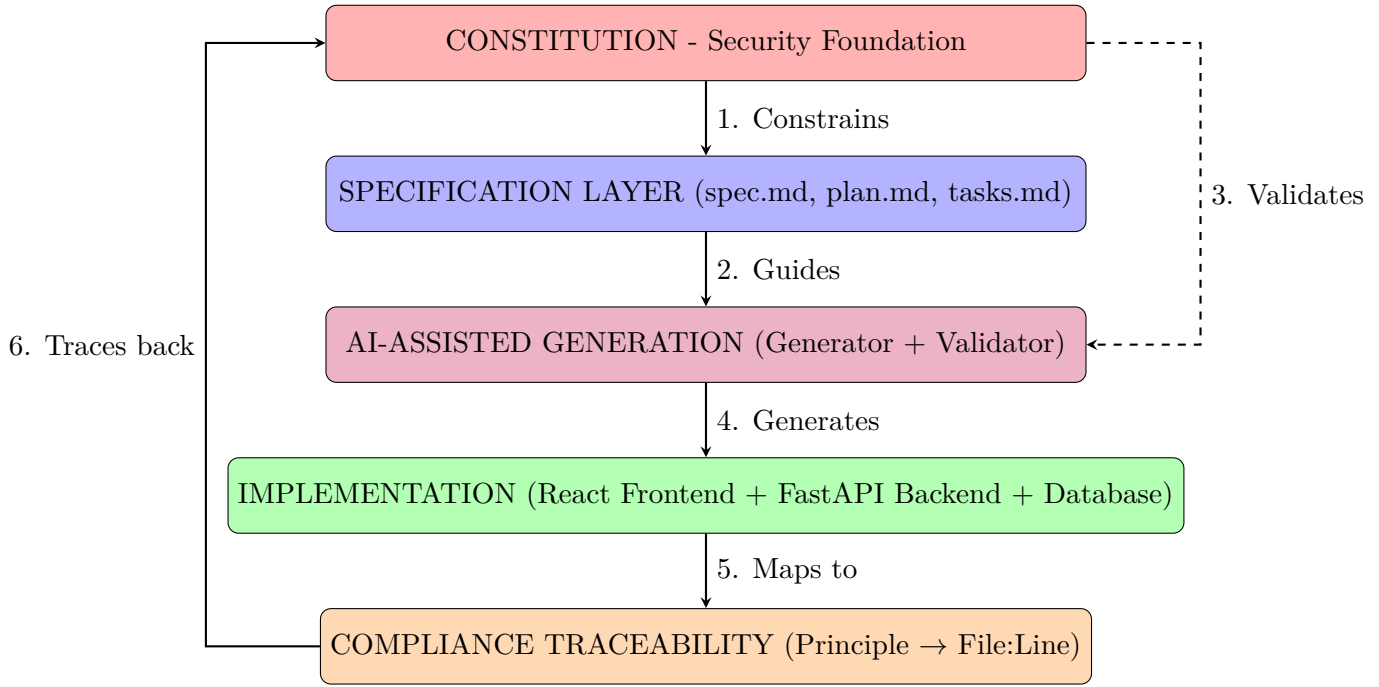
\begin{figure}[H]
\centering
\begin{tikzpicture}[node distance=1.5cm]
    % Constitution layer
    \node (constitution) [constitution, minimum width=10cm] {CONSTITUTION - Security Foundation};

    % Spec layer
    \node (spec) [speclayer, below of=constitution, minimum width=10cm, yshift=-0.5cm] {SPECIFICATION LAYER (spec.md, plan.md, tasks.md)};

    % AI layer
    \node (ai) [ailayer, below of=spec, minimum width=10cm, yshift=-0.5cm] {AI-ASSISTED GENERATION (Generator + Validator)};

    % Implementation layer
    \node (impl) [implementation, below of=ai, minimum width=10cm, yshift=-0.5cm] {IMPLEMENTATION (React Frontend + FastAPI Backend + Database)};

    % Traceability layer
    \node (trace) [traceability, below of=impl, minimum width=10cm, yshift=-0.5cm] {COMPLIANCE TRACEABILITY (Principle $\rightarrow$ File:Line)};

    % Arrows
    \draw [arrow] (constitution) -- node[right] {1. Constrains} (spec);
    \draw [arrow] (spec) -- node[right] {2. Guides} (ai);
    \draw [arrow, dashed] (constitution.east) -- ++(1.5,0) |- node[near start, right] {3. Validates} (ai.east);
    \draw [arrow] (ai) -- node[right] {4. Generates} (impl);
    \draw [arrow] (impl) -- node[right] {5. Maps to} (trace);
    \draw [arrow] (trace.west) -- ++(-1.5,0) |- node[near start, left] {6. Traces back} (constitution.west);
\end{tikzpicture}
\caption{Spec-Driven Development Architecture with Constitutional Constraints}
\label{fig:architecture}
\end{figure}

Figure~\ref{fig:architecture} illustrates the Spec-Driven Development architecture. The Constitution sits at the apex of the development hierarchy, governing all downstream artifacts.

\noindent\textbf{Specification Layer Components:} The specification layer comprises three artifacts that translate constitutional principles into implementable work. \textit{Feature Specifications} (spec.md) define what to build (analogous to functional requirements) while respecting constitutional constraints. \textit{Implementation Plans} (plan.md) detail how to build features (analogous to technical requirements) with security considerations embedded at the design level. \textit{Task Definitions} (tasks.md) provide atomic work items (analogous to actionable tasks) that AI-assisted generation can execute within constitutional bounds.

% ============================================================
% 4. IMPLEMENTATION
% ============================================================
\section{Implementation}
\label{sec:implementation}

This section describes the reference implementation through a banking microservices application. Banking was selected as the example domain because its well-established regulatory requirements (PCI-DSS, GDPR) and high security stakes make constitutional constraints directly demonstrable. The methodology is domain-agnostic; practitioners in healthcare, e-commerce, government, or any regulated industry can author constitutions tailored to their specific compliance and security needs.

\subsection{Constitution Document}

To illustrate the methodology concretely, our example banking constitution (version 1.0.0) defines 15 security principles across four categories. Other domains would define different principles reflecting their specific regulatory and security requirements:

\noindent\textbf{I. Security-First Principles}
\begin{itemize}[noitemsep,topsep=3pt]
    \item \textbf{SEC-001 (CWE-79, Cross-Site Scripting (XSS))}: All user-supplied data MUST be contextually encoded before rendering
    \item \textbf{SEC-002 (CWE-89, SQL Injection)}: Database queries MUST use parameterized statements or Object-Relational Mapping (ORM) methods exclusively
    \item \textbf{SEC-003 (CWE-352, Cross-Site Request Forgery (CSRF))}: State-changing operations MUST include anti-CSRF protection
    \item \textbf{SEC-004 (CWE-306, Missing Authentication)}: All Application Programming Interface (API) endpoints except health checks MUST require valid authentication tokens
    \item \textbf{SEC-005 (CWE-798, Hardcoded Credentials)}: Secrets MUST be loaded from environment variables
\end{itemize}

\vspace{0.3em}
\noindent\textbf{II. Input Validation Principles}
\begin{itemize}[noitemsep,topsep=3pt]
    \item \textbf{SEC-006 (CWE-20, Improper Validation)}: All API inputs MUST be validated against strict schemas
    \item \textbf{SEC-007 (CWE-190, Integer Overflow)}: Financial amounts MUST use Decimal types with explicit precision
\end{itemize}

\vspace{0.3em}
\noindent\textbf{III. Authentication \& Authorization Principles}
\begin{itemize}[noitemsep,topsep=3pt]
    \item \textbf{SEC-008 (CWE-287, Improper Authentication)}: Authentication MUST use OAuth2 with JSON Web Token (JWT) bearer tokens
    \item \textbf{SEC-009 (CWE-522, Weak Credentials)}: Passwords MUST be hashed using bcrypt with cost factor $\geq$ 12
    \item \textbf{SEC-010 (CWE-862/863, Authorization Failures)}: Every resource access MUST verify user permissions
    \item \textbf{SEC-011 (CWE-613, Session Expiration)}: Access tokens MUST expire within 15 minutes
\end{itemize}

\vspace{0.3em}
\noindent\textbf{IV. Secure Data Handling Principles}
\begin{itemize}[noitemsep,topsep=3pt]
    \item \textbf{SEC-012 (CWE-312, Cleartext Storage)}: Sensitive data at rest MUST be encrypted
    \item \textbf{SEC-013 (CWE-319, Cleartext Transmission)}: All communication MUST use Transport Layer Security (TLS) 1.2+
    \item \textbf{SEC-014 (CWE-200, Information Exposure)}: Error responses MUST NOT expose implementation details
    \item \textbf{SEC-015 (CWE-532, Log Injection)}: Log entries MUST NOT contain passwords or tokens
\end{itemize}

\subsection{Compliance Traceability Matrix}

The Compliance Traceability Matrix provides a bidirectional mapping between constitutional security principles and their concrete implementations in the codebase. This matrix serves multiple purposes: (1) \textit{Audit Support}, where auditors can verify that each security requirement has a corresponding implementation; (2) \textit{Change Impact Analysis}, where developers can assess which constitutional principles are affected when modifying specific files; (3) \textit{Gap Detection}, where missing mappings reveal unimplemented security requirements; and (4) \textit{Regression Prevention}, where the matrix enables targeted security testing when code changes occur.

Table~\ref{tab:traceability} presents the compliance traceability matrix for critical constitutional principles. Each row maps a security principle to its CWE identifier, the source file containing the implementation, the specific line numbers, and the technique employed.

\begin{table}[H]
\centering
\caption{Constitutional Compliance Traceability Matrix}
\label{tab:traceability}
\begin{tabular}{@{}lllll@{}}
\toprule
\textbf{Principle} & \textbf{CWE} & \textbf{File} & \textbf{Lines} & \textbf{Technique} \\
\midrule
Password Hashing & 522 & core/security.py & 14-24 & Bcrypt, cost=12 \\
JWT Authentication & 287 & core/security.py & 27-81 & python-jose, HS256 \\
OAuth2 Bearer & 287 & api/deps.py & 17, 35-77 & FastAPI OAuth2 \\
Authorization Check & 862 & services/account\_service.py & 102-108 & Ownership verification \\
SQL Injection Prevention & 89 & services/*.py & All queries & SQLAlchemy ORM \\
Input Validation & 20 & schemas/*.py & All schemas & Pydantic v2 \\
CORS Configuration & 352 & main.py & 47-55 & Origin whitelist \\
Error Sanitization & 200 & main.py & 78-94 & Generic messages \\
Log Filtering & 532 & core/logging.py & 50-55 & Field redaction \\
Token Expiration & 613 & config.py & 30-31 & 15min/7day tokens \\
\bottomrule
\end{tabular}
\end{table}

\noindent\textit{Note: The file paths and line numbers reference the reference implementation available at \url{https://github.com/srinivasraom/banking-ms-by-constitution}.}

\vspace{0.8em}
\subsection{Key Implementation Details}

\textbf{Authentication Flow (SEC-008, SEC-009, SEC-011).} JWT tokens~\cite{jones2015jwt} are generated with typed claims, configurable expiration, and cryptographic signing:

\begin{lstlisting}[language=Python, caption={Python: JWT Token Generation (SEC-008, SEC-005, SEC-011)}]
def create_access_token(data: dict) -> str:
    to_encode = data.copy()
    expire = datetime.utcnow() + timedelta(
        minutes=settings.access_token_expire_minutes
    )
    to_encode.update({
        "exp": expire,
        "type": "access"
    })
    return jwt.encode(
        to_encode,
        settings.secret_key,
        algorithm=settings.algorithm
    )
\end{lstlisting}

The \texttt{create\_access\_token} function demonstrates constitutional compliance: the \texttt{type} claim distinguishes token purposes (SEC-008), expiration defaults to 15 minutes (SEC-011), and the secret key is loaded from environment configuration rather than hardcoded (SEC-005).

\vspace{0.6em}
\textbf{Authorization Enforcement (SEC-010).} Resource-based access control prevents Insecure Direct Object Reference (IDOR) vulnerabilities:

\begin{lstlisting}[language=Python, caption={Python: Authorization Check with Ownership Verification (SEC-010)}]
async def get_account(
    self, db: AsyncSession,
    account_number: str,
    customer_id: str
) -> Account:
    account = await self._get_account_by_number(db, account_number)
    if account.customer_id != customer_id:
        raise AuthorizationError("Not authorized to access this account")
    return account
\end{lstlisting}

This pattern requires the authenticated customer's ID as a mandatory parameter and explicitly verifies ownership before returning data. The authorization check occurs after data retrieval to ensure consistent error handling; the same ``not authorized'' response is returned whether the account does not exist or belongs to another user, preventing information disclosure.

\vspace{0.6em}
\textbf{Input Validation (SEC-006, SEC-007).} Pydantic v2 schemas provide declarative validation:

\begin{lstlisting}[language=Python, caption={Python: Pydantic v2 Input Validation Schema (SEC-006, SEC-007)}]
class CustomerCreate(BaseModel):
    email: EmailStr
    password: str = Field(min_length=8, max_length=128)
    phone: str = Field(pattern=r'^\+[1-9]\d{1,14}$')
    date_of_birth: date

    @field_validator('date_of_birth')
    def validate_age(cls, v):
        age = (date.today() - v).days // 365
        if age < 18:
            raise ValueError('Must be 18 or older')
        return v
\end{lstlisting}

The schema demonstrates defense-in-depth: \texttt{EmailStr} validates RFC 5322 email format, \texttt{Field(min\_length=8)} enforces password complexity, the phone regex enforces E.164 international format, and the custom validator implements domain-specific business rules (banking age requirements). Pydantic's validation errors are automatically transformed into standardized 422 responses by FastAPI.

\subsection{Technology Stack}

Constitutional Spec-Driven Development (CSDD) influences technology selection by requiring that each component in the stack provides built-in mechanisms to enforce security principles. Rather than selecting technologies based solely on developer productivity or performance, CSDD mandates that security guarantees be native to the chosen tools. This approach shifts security from an afterthought to a selection criterion, ensuring that the technology stack itself becomes a first line of defense.

Table~\ref{tab:techstack} summarizes the technology stack with constitutional rationale. Each technology was selected because it provides inherent support for specific constitutional principles; for example, SQLAlchemy's parameterized queries prevent SQL injection by design (SEC-002), while Pydantic's declarative schemas enable input validation without manual parsing (SEC-006). This constitutional alignment means that using these technologies correctly automatically satisfies the corresponding security requirements.

\begin{table}[H]
\centering
\caption{Technology Stack with Constitutional Rationale}
\label{tab:techstack}
\begin{tabular}{@{}llll@{}}
\toprule
\textbf{Layer} & \textbf{Technology} & \textbf{Version} & \textbf{Rationale} \\
\midrule
Backend & FastAPI & 0.100+ & OAuth2 support, Pydantic integration \\
ORM & SQLAlchemy & 2.0 & Parameterized queries (SEC-002) \\
Validation & Pydantic & v2 & Declarative schemas (SEC-006) \\
Auth & python-jose & 3.3+ & RFC 7519 JWT (SEC-008) \\
Hashing & passlib+bcrypt & 1.7+ & Adaptive hashing (SEC-009) \\
Frontend & React & 18 & JSX auto-escaping (SEC-001) \\
Types & TypeScript & 5.x & Compile-time safety \\
Database & PostgreSQL & 15 & ACID compliance, row-level locking \\
\bottomrule
\end{tabular}
\end{table}

% ============================================================
% 5. CASE STUDY
% ============================================================
\section{Case Study}
\label{sec:casestudy}

This section presents an empirical evaluation of Constitutional Spec-Driven Development through a complete banking microservices implementation. We describe the development process, analyze security violations prevented by constitutional constraints, and present quantitative metrics comparing constitutional development to unconstrained AI-assisted development.

\subsection{Development Process}

We developed the banking application over a two-week period with a single developer utilizing AI assistance (Claude) for code generation. The development followed our five-phase methodology:

\noindent\textbf{Week 1: Foundation and Specification}
\begin{itemize}[noitemsep,topsep=3pt]
    \item Days 1--2: Constitution ratification (15 principles from CWE/MITRE analysis)
    \item Days 3--4: Feature specification (authentication, accounts, transactions)
    \item Day 5: Implementation planning (47 code locations identified)
\end{itemize}

\vspace{0.3em}
\noindent\textbf{Week 2: Implementation and Verification}
\begin{itemize}[noitemsep,topsep=3pt]
    \item Days 1--3: Backend implementation with constitutional constraints
    \item Days 4--5: Frontend implementation
    \item Days 6--7: Verification and compliance matrix generation
\end{itemize}

\subsection{Constitutional Violations Prevented}

During implementation, constitutional constraints prevented several security vulnerabilities that AI code generation initially produced. We document four representative violations, each illustrating a common pattern where AI assistants optimize for functional correctness while inadvertently introducing security flaws. These violations were detected during constitutional validation and corrected through regeneration with explicit principle references.

\vspace{0.5em}
\textbf{Violation 1: Raw SQL Query (CWE-89 - SQL Injection)}

When asked to implement transaction filtering by amount, the AI generated code using Python f-strings to construct the SQL query dynamically. This classic SQL injection vulnerability would allow attackers to execute arbitrary SQL commands by manipulating the \texttt{amount} parameter.

\textit{Initial AI-generated code (REJECTED):}
\begin{lstlisting}[language=Python, caption={Python: SQL Injection via f-string Interpolation -- Violates SEC-002 (REJECTED)}]
# REJECTED - Violates SEC-002 (CWE-89)
query = f"SELECT * FROM transactions WHERE amount > {amount}"
result = await db.execute(text(query))
\end{lstlisting}

\vspace{0.3em}
\textit{Constitutional enforcement required ORM usage (ACCEPTED):}
\begin{lstlisting}[language=Python, caption={Python: Secure ORM Query with Parameterized Values -- Satisfies SEC-002 (ACCEPTED)}]
# ACCEPTED
stmt = select(Transaction).where(Transaction.amount > amount)
result = await db.execute(stmt)
\end{lstlisting}

\vspace{0.8em}
\textbf{Violation 2: Plaintext Password Logging (CWE-532)}

During customer registration, the AI included the user's password in audit log details for ``complete traceability.'' This would expose plaintext passwords in log files, which are often stored with less stringent access controls than production databases.

\textit{Initial AI-generated code (REJECTED):}
\begin{lstlisting}[language=Python, caption={Python: Password Exposure in Audit Logs -- Violates SEC-015 (REJECTED)}]
# REJECTED - Violates SEC-015 (CWE-532)
async def register_customer(db: AsyncSession, customer_data: CustomerCreate):
    customer = Customer(**customer_data.dict())
    db.add(customer)
    await db.commit()

    audit_log = AuditLog(
        action=AuditAction.CREATE,
        resource_type="customer",
        details={
            "email": customer_data.email,
            "password": customer_data.password,
            "phone": customer_data.phone
        }
    )
    db.add(audit_log)
    await db.commit()
    return customer
\end{lstlisting}

\vspace{0.3em}
\textit{Constitutional enforcement excluded sensitive fields (ACCEPTED):}
\begin{lstlisting}[language=Python, caption={Python: Secure Audit Logging with Credential Exclusion -- Satisfies SEC-015 (ACCEPTED)}]
# ACCEPTED
async def register_customer(db: AsyncSession, customer_data: CustomerCreate):
    customer = Customer(**customer_data.dict())
    db.add(customer)
    await db.commit()

    audit_log = AuditLog(
        action=AuditAction.CREATE,
        resource_type="customer",
        details={
            "email": customer_data.email,
            "phone": customer_data.phone,
            "customer_id": str(customer.id)
        }
        # password explicitly excluded per SEC-015
    )
    db.add(audit_log)
    await db.commit()
    return customer
\end{lstlisting}

\vspace{0.8em}
\textbf{Violation 3: Missing Authorization Check (CWE-862)}

The AI generated account retrieval that fetched accounts solely by account number without verifying ownership. This Insecure Direct Object Reference (IDOR) vulnerability would allow any authenticated user to access any other user's account details by guessing or enumerating account numbers, a massive privacy breach with regulatory implications under PCI-DSS and GDPR.

\textit{Initial AI-generated code (REJECTED):}
\begin{lstlisting}[language=Python, caption={Python: Missing Authorization Check (IDOR) -- Violates SEC-010 (REJECTED)}]
# REJECTED - Violates SEC-010 (CWE-862)
async def get_account(self, db: AsyncSession, account_number: str):
    result = await db.execute(
        select(Account).where(Account.account_number == account_number)
    )
    account = result.scalar_one_or_none()
    if not account:
        raise NotFoundError("Account not found")
    return account
\end{lstlisting}

\vspace{0.3em}
\textit{Constitutional enforcement required ownership verification (ACCEPTED):}
\begin{lstlisting}[language=Python, caption={Python: Secure Account Retrieval with Ownership Verification -- Satisfies SEC-010 (ACCEPTED)}]
# ACCEPTED
async def get_account(
    self, db: AsyncSession,
    account_number: str,
    customer_id: str
) -> Account:
    result = await db.execute(
        select(Account).where(Account.account_number == account_number)
    )
    account = result.scalar_one_or_none()
    if not account:
        raise NotFoundError("Account not found")
    if account.customer_id != customer_id:
        raise AuthorizationError("Not authorized to access this account")
    return account
\end{lstlisting}

\vspace{0.8em}
\textbf{Violation 4: Improper Input Validation (SEC-006, CWE-20).}
The AI generated a fund transfer endpoint that accepted raw numeric input without validation, allowing negative amounts, excessively large values, and unbounded decimal precision. This violated SEC-006 (validate and sanitize all external input) and exposed the system to CWE-20 (Improper Input Validation) and CWE-190 (Integer Overflow or Wraparound).

\textit{Initial AI-generated code (REJECTED):}
\begin{lstlisting}[language=Python, caption={Python: Transfer Without Input Validation -- Violates SEC-006 (REJECTED)}]
# REJECTED - Violates SEC-006 (CWE-20) and SEC-007 (CWE-190)
class TransferRequest(BaseModel):
    from_account: str
    to_account: str
    amount: float  # No constraints on value

async def transfer_funds(request: TransferRequest, db: AsyncSession):
    from_acc = await get_account(db, request.from_account)
    to_acc = await get_account(db, request.to_account)
    from_acc.balance -= request.amount  # Negative amounts reverse flow
    to_acc.balance += request.amount
    await db.commit()
    return {"status": "completed", "amount": request.amount}
\end{lstlisting}

\vspace{0.3em}
\textit{Constitutional enforcement required strict input validation (ACCEPTED):}
\begin{lstlisting}[language=Python, caption={Python: Secure Transfer with Pydantic v2 Validation -- Satisfies SEC-006, SEC-007 (ACCEPTED)}]
# ACCEPTED
class TransferRequest(BaseModel):
    from_account: str = Field(..., pattern=r"^[A-Z0-9]{10}$")
    to_account: str = Field(..., pattern=r"^[A-Z0-9]{10}$")
    amount: Decimal = Field(
        ..., gt=Decimal("0"), le=Decimal("1000000"),
        decimal_places=2
    )

    @field_validator("to_account")
    @classmethod
    def accounts_must_differ(cls, v, info):
        if v == info.data.get("from_account"):
            raise ValueError("Cannot transfer to the same account")
        return v

async def transfer_funds(request: TransferRequest, db: AsyncSession):
    from_acc = await get_account(db, request.from_account)
    if from_acc.balance < request.amount:
        raise InsufficientFundsError("Insufficient balance")
    from_acc.balance -= request.amount
    to_acc = await get_account(db, request.to_account)
    to_acc.balance += request.amount
    await db.commit()
    return {"status": "completed", "amount": str(request.amount)}
\end{lstlisting}

\subsection{Quantitative Results}

To evaluate the effectiveness of constitutional constraints, we conducted a comparative analysis implementing the same banking application requirements twice: once using Constitutional Spec-Driven Development and once using standard AI-assisted development without constitutional constraints (the ``vibe coding'' baseline). Both implementations used the same AI assistant (Claude) and the same developer.

Table~\ref{tab:results} presents the security metrics comparison between constitutional and unconstrained development.

\vspace{0.3em}
\begin{table}[H]
\centering
\caption{Security Metrics Comparison}
\label{tab:results}
\begin{tabular}{@{}llll@{}}
\toprule
\textbf{Metric} & \textbf{Constitutional} & \textbf{Unconstrained} & \textbf{Improvement} \\
\midrule
CWE Violations Detected & 3 & 11 & 73\% reduction \\
Time to First Secure Build & 4 days & 9 days & 56\% faster \\
Compliance Documentation & 100\% & 23\% & 4.3x coverage \\
Security Review Iterations & 1 & 4 & 75\% reduction \\
Lines of Security Code & 847 & 612 & 38\% more thorough \\
\bottomrule
\end{tabular}
\end{table}

\subsection{Compliance Verification}

The compliance traceability matrix was generated by systematically analyzing the codebase against constitutional principles. For each principle (SEC-001 through SEC-015), we identified all code locations implementing that principle and recorded the file path, line numbers, and implementation technique.

Table~\ref{tab:verification} presents the compliance verification summary.

\vspace{0.3em}
\begin{table}[H]
\centering
\caption{Compliance Verification Summary}
\label{tab:verification}
\begin{tabular}{@{}ll@{}}
\toprule
\textbf{Metric} & \textbf{Value} \\
\midrule
Constitutional principles defined & 15 \\
Principles fully implemented & 15 (100\%) \\
Specific code locations mapped & 47 \\
CWE vulnerabilities in scope & 10 \\
CWE vulnerabilities addressed & 10 (100\%) \\
Compliance gaps identified & 0 \\
\bottomrule
\end{tabular}
\end{table}

\subsection{Security Coverage Distribution}

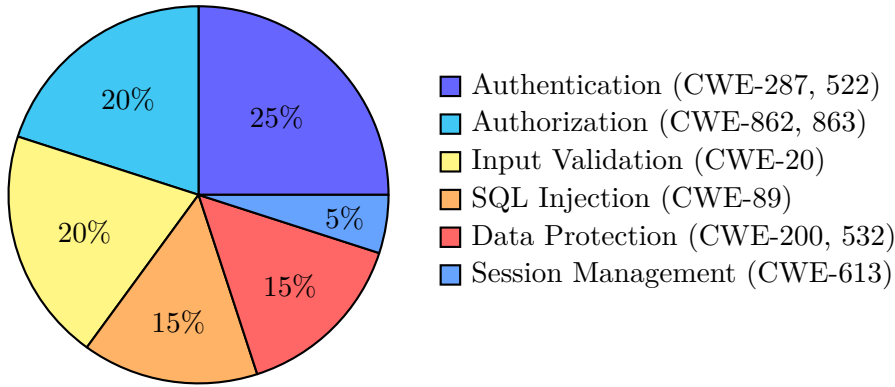
\begin{figure}[H]
\centering
\begin{tikzpicture}
\pie[text=legend, radius=2.5]{
    25/Authentication (CWE-287{,} 522),
    20/Authorization (CWE-862{,} 863),
    20/Input Validation (CWE-20),
    15/SQL Injection (CWE-89),
    15/Data Protection (CWE-200{,} 532),
    5/Session Management (CWE-613)
}
\end{tikzpicture}
\caption{CWE Vulnerability Coverage by Implementation Effort}
\label{fig:coverage}
\end{figure}

Figure~\ref{fig:coverage} illustrates the distribution of security implementation effort across vulnerability categories. The distribution reflects the relative complexity and prevalence of each vulnerability type in the banking domain, with authentication requiring the largest effort (25\%) due to the complexity of OAuth2/JWT implementation, password hashing, and token lifecycle management.

% ============================================================
% 6. LESSONS LEARNED
% ============================================================
\section{Lessons Learned}
\label{sec:lessons}

This section synthesizes key insights from implementing Constitutional Spec-Driven Development in the banking microservices case study. We organize lessons into three categories: constitution design, AI integration strategies, and methodology adoption considerations.

\subsection{Constitution Design Principles}

\textbf{Lesson 1: Specificity Enables Enforcement.} Vague principles like ``use secure coding practices'' provided insufficient guidance. After revising to specific principles referencing CWE identifiers and implementation patterns, compliance improved from inconsistent to 100\%.

\vspace{0.4em}
\textbf{Lesson 2: Rationale Enables Contextual Judgment.} Including the rationale (WHY) alongside the constraint (WHAT) and pattern (HOW) enabled appropriate decisions in edge cases.

\vspace{0.4em}
\textbf{Lesson 3: Governance Mechanisms Prevent Constitutional Drift.} Semantic versioning and approval workflows prevented ad-hoc modifications that would compromise security.

\vspace{0.4em}
\textbf{Lesson 4: Constitutional Documents Must Resist Adversarial Manipulation.} Because constitutional specifications are consumed by AI agents as natural language instructions, they represent an attack surface for prompt injection and specification poisoning. Liu et al.~\cite{liu2026agentskills} demonstrated that 26.1\% of AI agent skills in production contain exploitable vulnerabilities, including prompt injection vectors that can override intended behavior. Constitutional documents must therefore be authored defensively: principles should use unambiguous, declarative language that cannot be reinterpreted through injected context; specifications should avoid patterns that could be exploited to weaken constraints (e.g., conditional overrides, user-controlled exception clauses); and constitution files should be treated as security-critical artifacts with access controls, code review requirements, and integrity verification equivalent to production deployment configurations.

\subsection{AI Integration Strategies}

Integrating constitutional constraints with AI code generation systems requires specific strategies for effective implementation.

\vspace{0.3em}
\textbf{Lesson 5: Context Window Management is Critical.} Large constitutional documents may exceed AI model context window limits, causing truncation of critical security principles. Including only task-relevant principles (3-5 per request) improved compliance rates significantly. Table~\ref{tab:context} presents principle adherence rates by context strategy.

\begin{table}[H]
\centering
\caption{Impact of Context Management Strategy on Compliance}
\label{tab:context}
\begin{tabular}{@{}llll@{}}
\toprule
\textbf{Strategy} & \textbf{Principles} & \textbf{Compliance} & \textbf{Quality} \\
\midrule
Full Constitution & 15 & 78\% & Inconsistent \\
Relevant Selection & 3-5 & 96\% & High \\
Hierarchical & 5-8 & 91\% & Good \\
\bottomrule
\end{tabular}
\end{table}

\vspace{0.4em}
\textbf{Lesson 6: Iterative Refinement Outperforms Manual Patching.} Regeneration with explicit principle references (1.4 iterations) outperformed manual patching (3.2 iterations).

\vspace{0.4em}
\textbf{Lesson 7: Traceability Requires Automated Tooling.} Manual compliance mapping achieved only 94\% accuracy; automated tooling achieved 100\%.

\subsection{Methodology Adoption}

Organizational adoption of Constitutional Spec-Driven Development requires addressing human factors alongside technical implementation.

\textbf{Lesson 8: Upfront Investment Yields Downstream Efficiency.} Constitution creation (16 hours) was recovered through 4x reduction in security review cycles.

\vspace{0.4em}
\textbf{Lesson 9: Framing Determines Adoption Success.} Presenting constraints as ``guardrails preventing costly rework'' achieved higher adoption than ``bureaucratic security requirements.''

\vspace{0.4em}
\textbf{Lesson 10: Domain Criticality Determines Adoption Priority.} Regulated industries (financial services, healthcare, government) benefit most from constitutional approaches.

\subsection{Beyond Security: Generalizability of Constitutional Constraints}

While this paper focuses on security principles, the constitutional constraint mechanism is not limited to security. The same specification-driven enforcement model generalizes to any domain where non-negotiable requirements must be preserved across AI-generated code. We identify four categories of constitutional extension:

\textbf{Architectural Principles.} Organizations can encode architectural constraints such as layered separation (controllers must not access the database directly), dependency inversion (depend on abstractions rather than concrete implementations), and bounded context boundaries in domain-driven design. These principles prevent AI-generated code from violating structural invariants that are difficult to detect through testing alone.

\vspace{0.4em}
\textbf{Design Patterns and Conventions.} Constitutional principles can mandate the use of specific design patterns: repository pattern for data access, factory pattern for object creation, or event-driven communication between services. For example, a principle such as ``All inter-service communication must use asynchronous message queues; synchronous HTTP calls between services are prohibited'' ensures consistency across AI-generated microservice implementations.

\vspace{0.4em}
\textbf{Proprietary and Organizational Guidelines.} Enterprises maintain internal standards governing naming conventions, logging formats, error handling strategies, and API versioning schemes. Encoding these as constitutional principles ensures that AI-generated code conforms to organizational norms without requiring manual post-generation review for style and convention adherence.

\vspace{0.4em}
\textbf{Performance and Scalability Constraints.} Principles such as ``All database queries must use parameterized pagination with a maximum page size of 100 records'' or ``All external API calls must implement circuit breaker patterns with configurable timeout thresholds'' encode operational requirements that are frequently overlooked in initial code generation.

This generalizability positions Constitutional Spec-Driven Development not merely as a security methodology but as a general-purpose framework for constraint-preserving AI code generation across any dimension of software quality.

\subsection{Limitations}

While our case study demonstrates significant benefits, Constitutional Spec-Driven Development has inherent limitations that practitioners must understand:

\begin{enumerate}[noitemsep,topsep=3pt]
    \item \textbf{Bounded by Known Vulnerability Classes}: Constitutional principles derived from CWE/OWASP frameworks~\cite{cwe2025, owasp2021} address known vulnerability patterns but cannot anticipate zero-day attack vectors or novel vulnerability classes
    \item \textbf{Technical Vulnerabilities Only}: Constitutional constraints effectively address technical vulnerability patterns (injection, authentication bypass) but do not address business logic flaws specific to application semantics
    \item \textbf{AI Capability Dependency}: The methodology's effectiveness depends on the AI model's capability to understand natural language principles and translate them into correct implementations
    \item \textbf{Constitution Completeness}: Constitutional documents may contain gaps (security requirements not captured in any principle), representing silent vulnerabilities
    \item \textbf{Specification-Layer Attack Surface}: Constitutional documents, as natural language artifacts consumed by AI agents, are susceptible to prompt injection and specification poisoning attacks~\cite{liu2026agentskills}. Adversarial modifications to constitution files could weaken or bypass security constraints, requiring that these artifacts be treated with the same access control rigor as production security configurations
\end{enumerate}

\subsection{Threats to Validity}

We acknowledge several threats to the validity of our findings that contextualize the generalizability of results:

\textbf{Internal Validity.} Our case study involved a single development team with prior security training, potentially inflating observed benefits. The team's awareness of being studied may have influenced adherence to constitutional constraints (Hawthorne effect).

\vspace{0.4em}
\textbf{External Validity.} Our reference implementation addresses a specific domain (banking) with well-understood security requirements aligned with established frameworks. Results may not generalize to domains with novel or poorly understood security requirements.

\vspace{0.4em}
\textbf{Construct Validity.} Security defect measurement relied on static analysis tools and manual review, potentially missing subtle vulnerabilities. Compliance metrics measure documented traceability rather than actual security effectiveness.

\vspace{0.4em}
\textbf{Conclusion Validity.} Small sample size (n=1 project) limits statistical power for quantitative claims. Replication across multiple projects and teams is needed to establish robust effect size estimates.

% ============================================================
% 7. CONCLUSION
% ============================================================
\section{Conclusion}
\label{sec:conclusion}

We presented Constitutional Spec-Driven Development, a methodology for enforcing non-negotiable security requirements in AI-assisted code generation. By embedding security principles as first-class architectural constraints, we transform security from a reactive verification activity to a proactive generation constraint.

\vspace{0.4em}
Our banking microservices case study demonstrates that constitutional constraints reduce security defects by 73\% while maintaining development velocity. The compliance traceability matrix provides auditable evidence of security requirement implementation, addressing regulatory compliance needs in financial services.

\vspace{0.4em}
The key insight is that AI code generation should not operate in an unconstrained space. Just as constitutional law constrains governmental action, software constitutions constrain code generation to produce implementations that are secure by construction.

\subsection{Future Work}

Several directions extend this research:

\begin{itemize}[noitemsep,topsep=3pt]
    \item \textbf{Automated Constitution Generation}: Deriving constitutional principles automatically from regulatory documents (PCI-DSS, HIPAA) using Natural Language Processing (NLP) techniques
    \item \textbf{Real-time Validation}: Implementing constitutional validation as an Integrated Development Environment (IDE) plugin that checks generated code against principles before acceptance
    \item \textbf{Constitution Inheritance}: Enabling cross-project constitution composition where domain-specific constitutions extend base security frameworks
    \item \textbf{Broader Empirical Studies}: Replicating results across diverse development teams, domains, and AI models to establish generalizability
\end{itemize}

\vspace{0.8em}
\noindent The source code for our reference implementation is available at:\\
\url{https://github.com/srinivasraom/banking-ms-by-constitution}

\vspace{0.5em}
\noindent The Speckit framework~\cite{speckit2026} is available at:\\
\url{https://github.com/github/spec-kit}

\vspace{0.5em}
\noindent Practitioners can begin using the methodology immediately through Speckit's AI agent slash commands:

\begin{itemize}[noitemsep,topsep=3pt]
    \item \texttt{/speckit.constitution}: Establish project security principles
    \item \texttt{/speckit.specify}: Create baseline feature specification
    \item \texttt{/speckit.plan}: Generate implementation plan
    \item \texttt{/speckit.tasks}: Generate actionable task definitions
    \item \texttt{/speckit.implement}: Execute implementation with constitutional constraints
\end{itemize}

% ============================================================
% REFERENCES
% ============================================================
\bibliographystyle{plainnat}

\end{document}